\documentstyle[aps,multicol,epsf]{revtex}
\begin{document}

\newcommand{\rxy}{\rho_{xy}}
\newcommand{\rxx}{\rho_{xx}}
\newcommand{\sxy}{\sigma_{xy}}
\newcommand{\sxx}{\sigma_{xx}}
\newcommand{\ingaas}{In$_{0.53}$Ga$_{0.47}$As/InP}

\title{Evidence for a Quantum Hall Insulator in an
    InGaAs/InP Heterostructure}
\author{D.T.N. de Lang, L. A. Ponomarenko, A. de Visser}
\address{Van der Waals-Zeeman Instituut, Universiteit van Amsterdam,
Valckenierstraat 65, 1018 XE Amsterdam, The Netherlands}
\author{C. Possanzini, S.M. Olsthoorn}
\address{High Field Magnet Laboratory, University of
Nijmegen, 6525 ED Nijmegen, The Netherlands}
\author{A.M.M. Pruisken}
\address{Institute for Theoretical Physics, University of Amsterdam, 1018 XE Amsterdam, The Netherlands}
\date{\today} \maketitle

\begin{abstract} We study the quantum critical behavior of the plateau-insulator
(PI) transition in a low mobility \ingaas\ heterostructure. By
reversing the direction of the magnetic field ($B$) we find an
averaged Hall resistance $\rxy$ which remains quantized at the
plateau value $h/e^2$ throughout the PI transition. We extract a
critical exponent $\kappa '= 0.57 \pm 0.02$ for the PI transition
which is slightly different from (and possibly more accurate than)
the established value $0.42 \pm 0.04$ as previously obtained from
the plateau-plateau (PP) transitions.
\end{abstract}

\begin{multicols}{2}

One of the fundamental issues in the field of two dimensional
electron gases is the nature of the transitions between adjacent
quantum Hall plateaus. By measuring the resistance tensor of low
mobility \ingaas\ heterostructures, Wei {\it et
al.}~\cite{Wei,Hwang} demonstrated that the quantum Hall steps
become infinitely sharp as $T\rightarrow 0$, indicating that the
transitions between adjacent quantum Hall plateaus (PP
transitions) represent a sequence of quantum phase transitions
(QPT). Both the maximum slope in the Hall resistance with varying
$B$, $(\partial\rxy/\partial B)_{max}$, and the inverse of the
half-width of the longitudinal resistance between two adjacent
quantum Hall plateaus,  $(\Delta B)^{-1}$, have been shown to
follow the power law $T^{-\kappa}$ as $T$ approaches absolute
zero, independent of Landau level index. Here, $\kappa = p/2\nu$
where $p$ denotes the exponent of the phase breaking length
$\ell_\varphi$ at finite $T$ ({\it i.e.} $\ell_\varphi \sim
T^{-p/2}$) and $\nu$ is the critical index for the localization
length $\xi$ which is defined at zero $T$.

In order to probe the QPT, it is essential to carry out
experiments on samples where the dominant scattering mechanism is
provided by short ranged random potential
fluctuations~\cite{Wei2}. Like in \ingaas, this produces the
widest range in $T$ where quantum criticality is accessible
experimentally. At the same time, little is known about the
effects of macroscopic sample inhomogeneities which generally
complicate experiments on the QPT. The problem of sample
inhomogeneities was recently addressed by van Schaijk {\it et
al.}~\cite{Schaijk} who investigated the plateau-insulator (PI)
transition in the lowest Landau level. The data were taken from
the same \ingaas\ heterostructure which was previously used in the
study of the PP transitions~\cite{Hwang}.

Following the analysis by van Schaijk {\it et al.} one can extract
different exponents $\kappa$ and $\kappa '$ from the transport
data on the PI transition, dependent on the specific quantity one
considers. For example, the longitudinal resistance $\rxx$ was
shown to follow the exponential law~\cite{Shahar:SSC107}
$\rxx(\nu,T) \propto \exp(-\Delta\nu/\nu_0(T))$. Here, $\Delta\nu
= \nu - \nu_c$ represents the filling fraction $\nu$ of the lowest
Landau level relative to the critical value $\nu_c \approx
\frac{1}{2}$ and $\nu_0 (T) \propto T^{\kappa '}$ with an
experimental value $\kappa ' = 0.55 \pm 0.05$.

The numerical value of the exponent $\kappa '$ differs by more
than the experimental error from the established ``universal"
value $0.42 \pm 0.05$ that was previously extracted from the
resistance data on PP transitions~\cite{Wei,Hwang}. In an attempt
to understand the difference, van Schaijk {\it et al.} pointed out
that a different exponent ($\kappa \approx 0.42$) for the PI
transition is obtained by considering the temperature dependence
of the Hall conductance, $(\partial\sxy/\partial B)_{min} \propto
T^{-\kappa}$. It was shown that the different exponents $\kappa$
and $\kappa '$ are related to one another according to the
equation~\cite{Schaijk}

\begin{equation}
\label{kappacorr} \kappa=\kappa ' -
\frac{d\ln(\sxx^{*2}+1/4)}{d\ln T}  ,
\end{equation}
where $\sxx^*$ is the maximum value of $\sxx$ in units $e^2/h$.

Notice that for an ideal sample $\sxx^*$ is expected to be
universal and, hence, $\kappa$ and $\kappa '$ are identically the
same. In a real experiment, however, $\sxx^*$ usually depends
weakly on $T$. Moreover, a different absolute value is generally
found by performing different runs of the same experiment or by
reversing the field $B$. Eq.~\ref{kappacorr} therefore tells us
that the differences in the observed exponents $\kappa$ and
$\kappa '$ must be the result of macroscopic inhomogeneities in
the sample.

In this paper we further investigate the inhomogeneity problem and
extend the high $B$ results of van Schaijk {\it et
al.}~\cite{Schaijk} in several ways. We are specifically
interested in answering the question of universality of the
critical exponents, as well as the critical conductance $\sxx^*$.
For this purpose we study the effect of reversing the direction of
the $B$ field on the PP and PI transitions in general and on
Eq.~\ref{kappacorr} in particular.

Our sample and experimental setup are identical to those described
in Ref.~\cite{Schaijk}. The measurements were carried out in a
Bitter magnet ($B < 20$T) using a plastic dilution refrigerator.
The magnetotransport coefficients $\rxx$, $\rxy$ and the current
$I$ were measured simultaneously by using a standard AC technique
with a frequency of 13 Hz and an excitation current of 5 nA.

\begin{figure}[htb]
\epsfysize=19 \baselineskip
\centerline{\hbox{\epsffile{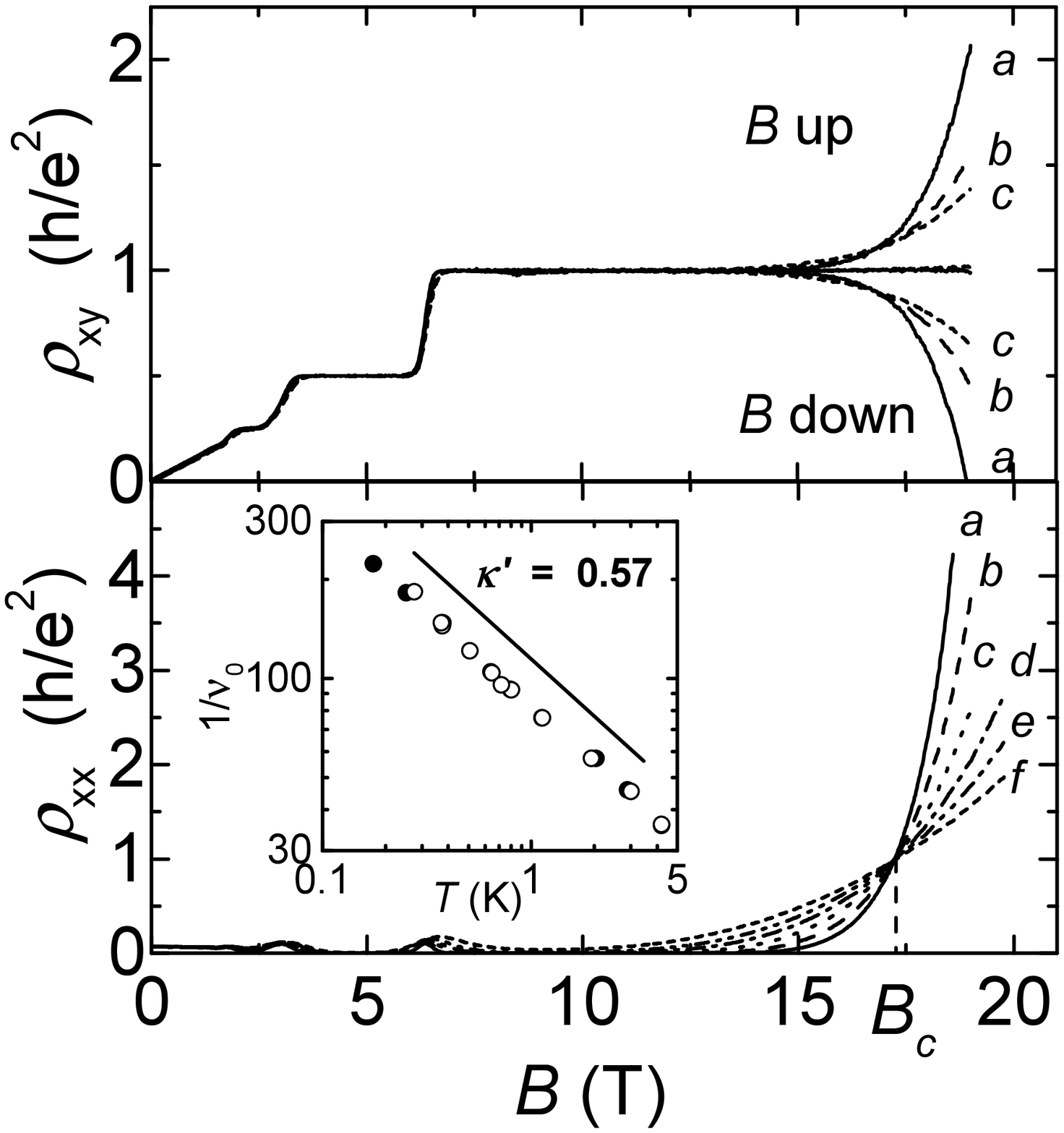}}} \caption{$\rxx$ and $\rxy$
 for a low mobility \ingaas\ heterojunction ($n = 2.2\times
10^{11}$cm$^{-2}, \mu = 16000$ cm$^2$/Vs) with varying $B$ for up
and down field directions. Labels $a,b,...f$ correspond to
temperatures 0.37, 0.62, 1.2, 1.9, 2.9 and 4.2 K. $B_c$ (= 17.2 T)
is the critical field for the plateau-insulator (PI) transition.
The $\rxx$ curves have been normalized to $\rxx(B_c)= h/e^2$.
Averaging over both field directions indicates that the Hall
resistance remains quantized beyond $B_c$. Inset: $1/\nu_0$ versus
$T$ for the PI transition indicating a critical exponent $\kappa '
= 0.57 \pm 0.02$. Closed symbols are data from positive field
directions, open symbols denote negative field directions. }
\end{figure}

Fig.~1 shows the results for sweeps in both directions of the $B$
field for different values of $T$. Upon reversing the direction of
$B$ at constant $T$ we find that the $\rxx$ data for the PI
transition remain unchanged. The $\rxy$ data, however, are
strongly affected and the results for opposite $B$ fields display
a symmetry about the plateau value $\rxy =  h/e^2$. By averaging
the $\rxy$ data over the $B$ directions we obtain a Hall
resistance that remains quantized also beyond the critical field
$B_c$ (= 17.2 T)  of the PI transition. This indicates that the
sequence of QPT's terminates in a so-called quantum Hall
insulating phase.

The phenomenon of a quantum Hall insulator has been observed on a
set of qualitatively different heterostructures and quantum wells,
such as Ge/SiGe~\cite{Hilke} and GaAs/AlGaAs~\cite{Shahar:SSC102}.
 However, these samples do not show any evidence for a QPT at the
lowest available $T$. In this case, the data on the PP and PI
transitions can be explained by semi-classical reasoning in
transport theory~\cite{Pruisken}.

The effect of reversing the $B$ field on the transport data of the
PI transition can qualitatively be understood as being the result
of a macroscopic misalignment of the Hall bar
contacts~\cite{Hilke,Furlan}. This kind of picture is in many ways
too simple, however, and it is more appropriate to think in terms
of macroscopic sample inhomogeneities such as electron density
fluctuations, inhomogeneous current distributions, etc. which
cannot be excluded from the experiment in general.

Notice that once the quantization of $\rxy$ throughout the PI
transition is accepted, the difference in $\kappa$ and $\kappa '$
is no longer an issue. The $\rxx$ data, with varying values of $T$
(Fig. 1), now display a true critical fixed point at the critical
field strength $B_c$. Therefore, contrary to van Schaijk {\it et
al.} , we must conclude that the critical index of the PI
transition is not given by $\kappa$ but, rather, by $\kappa '$
which is independent of the direction of $B$.

Next, we can make use of the renormalization theory of the quantum
Hall effect and remove the remaining experimental uncertainties in
the geometrical factor $L/W$ of the sample. Here, $L$ and $W$
stand for the length and width of the Hall bar respectively. This
factor is important since it determines the absolute value of
$\rxx$ and, hence, the correct value of the conductances $\sxx$
and $\sxy$.

An obvious criterion for fixing the value of $L/W$ is obtained by
demanding that the critical resistance $\rxx(B_c)$ be normalized
at $\rxx(B_c) = h/e^2$ such that the PI transition occurs
precisely at a half integral value of the Hall conductance, $\sxy
= \frac{1}{2}$, as it should be. The value of $L/W$, obtained in
this way, differs from a directly measured value by 8\% which is
quite reasonable.

As a result of the averaging procedure over the directions of $B$,
our data not only display particle-hole symmetry and scaling, but
also follow the statement of
duality~\cite{Shahar:SSC102,Kivelson}. It is important to remark
that the same averaging procedure has previously been studied in
detail for the PP transitions but the results are generally
somewhat different from those obtained in this paper. A detailed
study linking the PP and PI transitions will be published
elsewhere~\cite{deLang}.

As an important check on the consistency of our data, we next
extract the exponent $\kappa$ and the $\sxx^*$ (Eq.~1) for each
field direction independently. Fig. 2a shows the results for
$\kappa$ obtained from the slope of $\sxy$ for both directions of
the $B$ field. The insets show the data for $(\sxx^*)^2 + 1/4$
versus $T$ on a log-log scale. Following Eq.~\ref{kappacorr} we
obtain  $\kappa' = 0.56$, independently of the direction of $B$,
which agrees very well with the result $\kappa' = 0.57 \pm 0.02$
as obtained directly from the $T$ dependence of the $\rxx$ data.

The data from Ref.~\cite{Schaijk} are shown in Fig.~2b. Notice
that a different value for $\kappa$ was found but the results for
$\kappa '$ are the same as ours. Comparing the results of Figs~2a
and 2b we see that the value of $\kappa$ is generally different
for different field directions and different experimental runs.
The numerical value of $\kappa'$ remains constant in all cases,
however. This clearly indicates that $\kappa'$  represents the
true critical index.

\begin{figure}[htb]
\epsfysize=19 \baselineskip
\centerline{\hbox{\epsffile{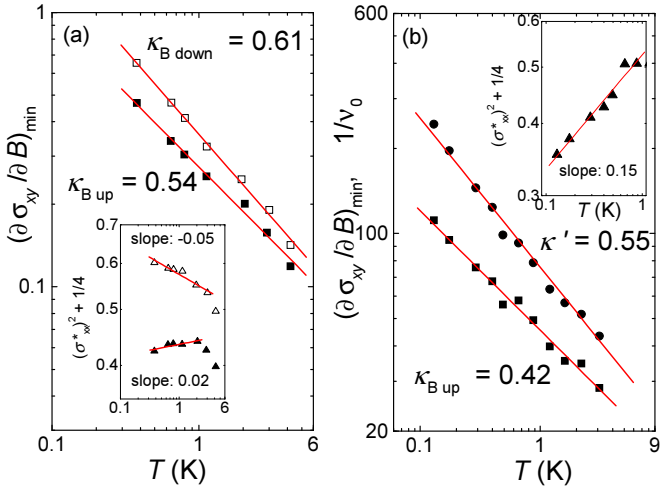}}}

\caption{(a) Field direction dependence of the critical exponent
$\kappa$ determined from the minimum slope of $\sxy$ with varying
$B$. Subtracting the correction factor according to Eq.~1 gives
the correct value of $\kappa ' = 0.57 \pm 0.02$ for both field
directions. (b)~Similarly, measurements by van Schaijk {\it et
al.}
give the same value of $\kappa '$ for the PI
transition. Closed symbols are positive B directions, while open
symbols denote negative fields. }
\end{figure}

In conclusion, we have measured and analyzed the PI transition of
a low mobility sample that exhibits scaling. The numerical value
of the critical exponent is estimated at $0.56 \pm 0.02$. This
upsets the ``established'' value of $0.42 \pm 0.04$ which was
extracted from the PP transitions and previously believed to be
universal. We attribute the different results for the PP and PI
transitions to the macroscopic sample inhomogeneities which have a
different effect on the qualitatively different transport data of
these transitions. However, more work on higher quality samples is
obviously necessary in order to narrow down the experimental
uncertainties in the numerical value of the critical indices.

\acknowledgments This research was in part supported by the Dutch
Science Foundation FOM and INTAS (Grant 99-1070).The authors would
like to thank H.P.~Wei for providing the sample.

\end{multicols}

\begin{references}
\bibitem{Wei}
H.P.~Wei, D.C.~Tsui, M.~Paalanen, and A.M.M.~Pruisken,
Phys.~Rev.~Lett.~{\bf 61}, 1294 (1988); A.M.M.~Pruisken,
Phys.~Rev.~Lett.~{\bf 61}, 1297 (1988).
\bibitem{Hwang}
S.W.~Hwang, H.P.~Wei, L.W.~Engel, D.C.~Tsui, and A.M.M.~Pruisken,
Phys.~Rev.~B~{\bf 48}, 11416 (1993).
\bibitem{Wei2}
H.P.~Wei , S.Y.~Lin, D.C.~Tsui, and A.M.M.~Pruisken,
Phys.~Rev.~B~{\bf 45}, 3926 (1992).
\bibitem{Schaijk}
R.T.F.~van~Schaijk,  A.~de~Visser, S.~Olsthoorn, H.P.~Wei, and
A.M.M.~Pruisken, Phys.~Rev.~Lett.~{\bf 84}, 1570 (2000).
\bibitem{Shahar:SSC107}
D.~Shahar, M.~Hilke, C.C.~Li, D.C.~Tsui, S.L.~Sondhi,
J.E.~Cunningham, and M.~Razeghi, Solid~State~Commun.~{\bf 107}, 19
(1998).
\bibitem{Hilke}
M.~Hilke, D.~Shahar, S.H.~Song, D.C.~Tsui, Y.H.~Xie, and
D.~Monroe, Nature~{\bf 395}, 675 (1998).
\bibitem{Shahar:SSC102}
D.~Shahar, D.C.~Tsui, M.~Shayegan, E.~Shimshoni, and S.L.~Sondhi,
Solid~State~Commun.~{\bf 102}, 817 (1997).
\bibitem{Pruisken}
A.M.M.~Pruisken, B.~\v{S}kori\'{c}, and M.A.~Baranov,
Phys.~Rev.~B~{\bf 60}, 16838 (1999).
\bibitem{Furlan}
M.~Furlan, Phys.~Rev.~B~{\bf 57}, 14818 (1998).
\bibitem{Kivelson}
S.~Kivelson, D.H.~Lee, and S.C.~Zhang, Phys.~Rev.~B~{\bf 46}, 2223
(1992).
\bibitem{deLang}
D.T.N.~de~Lang, L.~Ponomarenko, A.~de~Visser, and A.M.M.~Pruisken,
to be published.
\end{references}
\end{document}